\documentclass[aps,prd,reprint]{revtex4-1}
\usepackage{amsmath}
\usepackage{graphics}      
\usepackage{graphicx}      
\usepackage{longtable}     
\usepackage{url}           
\usepackage{bm}            
\usepackage[english]{babel}
\usepackage{pgfplots}
\usepackage{gensymb}

\begin{document}
\title{A counter-top search for macroscopic dark matter}
\author         {Jagjit Singh Sidhu, Glenn Starkman}
\affiliation    {Physics Department/CERCA/ISO Case Western Reserve University
Cleveland, Ohio 44106-7079, USA}
\author         {Ralph Harvey}
\affiliation    {Department of Earth, Environmental and Planetary Sciences
Case Western Reserve University
Cleveland, Ohio 44106-7079, USA}
\date{\today}

\begin{abstract}
A number of dark matter candidates have been discussed that are macroscopic, of approximately nuclear density, and scatter ordinary matter essentially elastically with approximately their geometric cross-section. 
A wide range of mass and geometric cross-section is still unprobed for these ``macros.''
Macros passing through rock would melt the material in cylinders surrounding their long nearly straight trajectories.  
Once cooled, the resolidified rock would be easily distinguishable from its surroundings.  
We discuss how, by visually examining ordinary slabs of rock such as are widely available commercially for kitchen countertops, 
one could probe an interesting segment of the open macro parameter space.
\end{abstract}
\maketitle
{\color{black}There is considerable, and widely known, 
evidence for the existence of dark matter \cite{Garrett2011}
or for the need to modify General Relativity 
(see \cite{Hinterbichler2012} for a pedagogical
review of modified gravity and 
\cite{deRham2017} for a more recent review
on the subject). }
Macroscopic dark matter (macros) represents a 
wide class of alternatives to particle dark matter --
large objects, probably composites of fundamental
particles, that are  ``dark'' because their
large mass implies a 
low number density and a small 
geometric cross-section per unit mass, 
even though the cross-section of each object is large.
There remains a large range of macro mass $M_X$ and geometric cross-section $\sigma_X$ that is still unprobed by experiments or observations.

A most intriguing possibility is that macros are made of Standard Model quarks or baryons bound by Standard Model forces.
This suggestion was originally
made by Witten \cite{PhysRevD.30.272},
in the context of a nuclear bag model and a then-possible first-order QCD phase transition.
A more realistic version was advanced by Lynn, Nelson and Tetradis
\cite{LYNN1990186} and Lynn again
\cite{1005.2124}, who argued 
in the context of $SU(3)$ chiral perturbation theory
that ``a bound state of baryons with a well-defined surface may conceivably form in the presence of kaon condensation.''
Nelson \cite{Nelson:1990iu} studied the possible
formation of such ``nuggets of strange baryon matter''
in an early-universe transition from a kaon-condensate phase of QCD to the ordinary phase.
Others have suggested non-Standard Model versions of such nuclear objects and their formation, for example incorporating the axion  \cite{Zhitnitsky:2002qa}.

\begin{figure*}
 \centering
       \includegraphics[width=\textwidth]{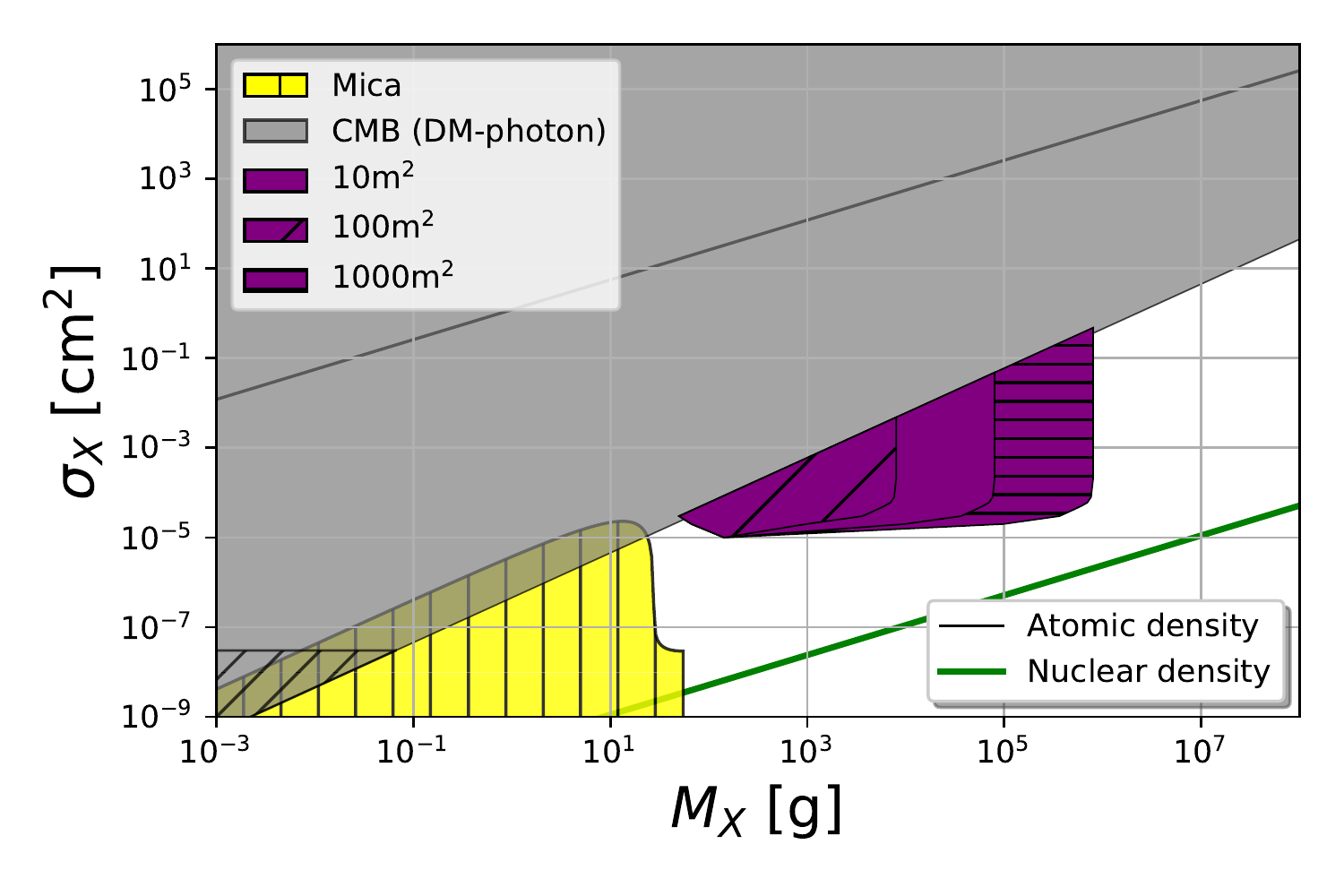}
        \caption{Illustrative regions of 
        parameter space that
        could be probed from observations
        of slabs of ordinary rock are shown in purple. 
        All three regions
        assume a minimum feature diameter of $1\,$mm.
	The region with diagonal hatching 
	assumes a total slab area of $10\,$m$^2$, 
	the unhatched region is for a slab of area of $100\,$m$^2$ and
	the horizontal-hatched region assumes a slab area of $1000\,$m$^2$. 
	The region currently excluded by examination of ancient mica \cite{DeRujula:1984axn, Price:1988ge} is shown in yellow with vertical hatching. 
	The grey region is excluded from the effects of cosmic microwave background photons scattering off the macros \cite{1309.7588}. The diagonal hatching
	of this bound in the bottom-left of the parameter space
	represents the region where the bound may weaken.
	Lines corresponding to nuclear and atomic density are shown for illustrative purposes.}
        \label{fig:rock}

 \end{figure*}

Such objects would presumably have densities that are comparable to nuclear density 
(which we take to be $\rho_{nuclear}=3.6 \times 10^{14}\,$g$\,$cm$^{-3}$).
This is much higher than ordinary ``atomic density'' 
($\rho_{atomic}=1\,$g$\,$cm$^{-3}$), 
and much lower density than black holes.  
The unconstrained macro parameter space includes a 
a wide range of $M_X$ for macros of nuclear density
(see \cite{jacobs2015macro} and 
\cite{jacobs2015resonant}). 
For $M_X\leq55\,$g, careful examination of specimens of old mica for tracks made by 
passing dark matter \cite{DeRujula:1984axn, Price:1988ge} have ruled out such objects 
as the primary dark matter candidate (see Figure \ref{fig:rock}).
For $M_X\geq10^{24}\,$g
, a variety of microlensing searches have similarly constrained 
macros \cite{Griest2013,Alcock2001,astro-ph/0607207,0912.5297}. 
For $M_X\gtrsim10^{15}\,$g, macros incident on white dwarfs would trigger thermonuclear runaways 
\cite{1505.04444}, as previously shown for primordial black holes \cite{1805.07381}, and are ruled out.
Reference \cite{1309.7588} utilized the full Boltzmann formalism to obtain constraints
from macro-photon elastic scattering using the first-year release of cosmic microwave background data from the Planck satellite.
{\color{black} 
Although strictly we should differentiate between the
macro-baryon cross-section and the macro-photon interaction,
these are macroscopic cross-sections and it is
difficult to imagine that these will differ significantly 
unless the photon wavelength is at least comparable to the size of the macro.
We therefore take 
the constraint in \cite{1309.7588} on $\sigma_{macro-photon}$
to apply to  $\sigma_x$ directly,
for $\sqrt{\frac{\sigma_x}{\pi}}\geq \lambda_{CMB} \sim 1000\,$nm (for photons at recombination).
For $\sqrt{\frac{\sigma_x}{\pi}}\leq \lambda_{CMB}$,
wave optics or quantum-mechanical effects may weaken
the constraints in a  model-dependent way.
We hatch the corresponding region of parameter space to
highlight this effect.}
Prior work had already constrained a similar range of 
parameter space by showing that
the consequence of dark-matter interactions with Standard Model particles is
to dampen the primordial matter fluctuations and 
suppress all structures below a given scale (see e.g. \cite{Bhm2001}).
In between, $55~\mathrm{g}\leq M_X \lesssim10^{15}~\mathrm{g}$, the coast is so far clear for nuclear-density macro dark matter. 

Recently, together with collaborators, we 
suggested how ultra-high-energy cosmic ray detectors that exploit atmospheric fluoresence could potentially be modified to probe parts of  macro parameter space \cite{Sidhu:2018auv}, including  macros of nuclear density and intermediate mass.
In this manuscript, we suggest how the approach 
applied to mica \cite{DeRujula:1984axn, Price:1988ge}
could be applied to a larger, widely available  sample of appropriate rock,
and used to search for larger-mass macros.
(``Paleo-detectors'' have also been considered recently for WIMPS \cite{Drukier:2018pdy,Edwards:2019puy}, {\color{black}
and earlier \cite{SnowdenIfft1995} using ancient mica.})
We also discuss 
our planned efforts to look for these
tracks directly and the range of macro mass and 
cross-section that could be probed.
For these purposes, we consider macros of a single mass and size,
even though a broad mass distribution is the expectation in the context of a composite dark matter candidate. 


The energy deposited by a transiting macro 
through its elastic scattering off the rock
is 
\begin{equation}
\frac{dE}{dx} = \sigma_X \rho v_X^2,
\end{equation}
where $\rho$ is the density of the 
medium, $\sigma_X$ is the geometric
cross section of the macro and
$v_X$ is the macro speed.

For definiteness, we assume macros possess a Maxwellian velocity distribution
\begin{equation}
	\label{eq:maxwellian}
	f_{MB}(v_X) = 
		\left( \frac{1}{\pi v_{vir}^2}\right)^{\frac{3}{2}}
		4\pi v_X^2 e^{-\left(\frac{v_X}{v_{vir}}\right)^2}, 
\end{equation}
where $v_{vir} \approx 250~ \,$km$\,$s$^{-1}$.  
This distribution is slightly modified by the Earth's motion.
(See footnote on page 15 of \cite{Sidhu:2018auv} for more details.) 
The cumulative velocity distribution function is obtained by integrating $f_{MB}$ up to the desired value of $v_X$. 
This allows us to determine the maximum mass $M_X$ we can probe as a function of $v_X$.

The speed of a macro traveling through
a medium is expected to evolve as
\begin{equation}
v = v_{X,0} e^{-\langle \rho x\rangle \frac{\sigma_X}{M_X}},
\end{equation}
where $\langle \rho x\rangle$ is
the encountered column density.
This will determine
the maximum value of $\frac{\sigma_X}{M_X}$ expected to
deposit sufficient energy
to produce an observable signal.

If a macro were to pass through  rock, 
the region nearest the trajectory would be
ionized, the surrounding region vaporized, and 
an even larger region would be melted.
The ionization and vaporization would result
in extreme pressures, especially near the trajectory \cite{1610.09680}. 
After resolidifying and cooling, the resulting 
rock would typically be petrologically distinguishable from the original rock around it.
For example, even much lower energy-density lightning strikes on sand rich in silica or quartz are known to form fulgurites\cite{pasek_block_pasek_2012}, 
glass tubes or clumps embedded in the sand.
If the macro were large enough,
the metamorphosed rock might even be visually distinguishable.
For example, when light-colored granite is melted,
it cools to forms a dark obsidian-like stone.
\footnote{
The major minerals of a typical light-colored granite are feldspar and quartz,  and 
the absorptions of these minerals are dominated by just a few specific bonds that are 
spread out across the spectrum.  Thus, most granites stay light-colored and usually 
whitish because the absorptions are spread around, although reddish and pinkish are 
not uncommon where there are more absorptions for some feldspars at the blue end of 
the spectrum.
	 Granitic rocks form from melts where considerable segregation of more easily-crystallized minerals has occured. They are therefore relatively rich in incompatible elements from across the periodic table. When melted and cooled too quickly to allow different phases to segregate, the resulting amorphous solid has absorptions all across the visible spectrum \cite{darkgranite,seconddarkgranite}.  Add in short-range order within the glass and one obtains a very dark material.
	 For example,
references \cite{darkgranite,seconddarkgranite} quote reflectivity
of
several percent.} 
	 
We solve the heat equation  to find 
out to what distance a  macro 
melts the rock surrounding its trajectory. 
This melt zone will cool to form a robust 
fossil record of the macro's passage.
Unlike fulgurites, the track of a macro would be 
straight, and, unless the properties of the rock are highly anisotropic, the trail is likely to consist of a highly distinctive, long straight cylinder of circular cross section.
This presents, in theory, a 
straightforward way of looking for 
macros in layers of rock.

Following the work of Cyncynates {\it et al.} 
\cite{1610.09680}, we approximate the 
initial heat deposition as a delta-function source along a straight line
through the rock, 
and propagate the heat outwards according to the heat 
equation. 
The resulting time-dependent temperature field 
\begin{equation}\label{tempfield}
T_c(r,t) = \frac{\sigma_{X} v_X^2}{4\pi \alpha c_p}\frac{e^{-\frac{r^2}{4t\alpha}}}{t}\,,
\end{equation}
where  $\sigma_X$ is the macro cross section,
$v_X$ is the macro speed,
$c_p$ is the specific heat capacity of the rock,
{\color{black} which we take to be $c_p \sim 900$ J kg$^{-1}$ K$^{-1}$ \cite{Waples2004}},
and $\alpha$ is its thermal diffusivity.

We invert \eqref{tempfield} to obtain 
$\pi r(t,T_c)^2$, the area that 
gets heated to $T_c$ (we take $T_c \approx 1200 
\degree C$\cite{melting}), such 
as the melting or vaporization 
temperature of the rock,
\begin{equation}\label{ionizationcrosssection}
\pi r(t,T_c)^2 = 4\pi t\alpha
	\ln(\frac{\sigma_{X} v_X^2}{4\pi \alpha c_p T_c t})\,.
\end{equation}

\begin{figure*}
 \centering
       \includegraphics[width=\textwidth]{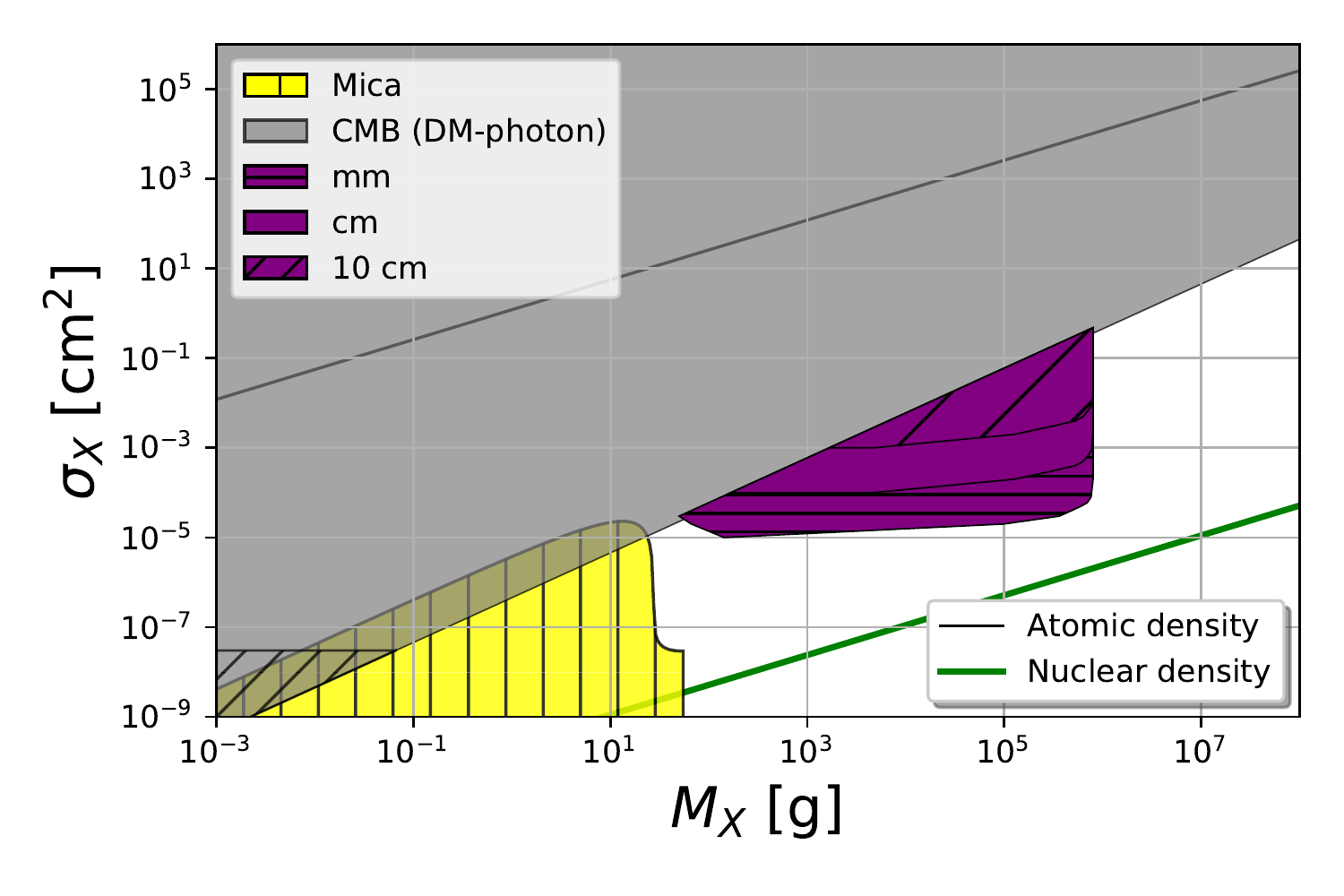}
        \caption{Illustrative regions of 
        parameter space that
        could be probed from observations
        of slabs of ordinary rock are shown in purple. 
        All 3 regions assume a slab area of $1000\,$m$^2$.
	The region with diagonal hatching 
	assumes
	searches for features with diameter of $0.1\,$cm or more, 
	the unhatched region is for a minimum feature diameter of $1\,$cm, and
	the horizontal-hatched region assumes a minimum feature diameter of $10\,$cm.
	The region currently excluded by examination of ancient mica \cite{DeRujula:1984axn, Price:1988ge} is shown in yellow with vertical hatching. 
	The grey region is excluded from the effects of cosmic microwave background photons scattering off the macros \cite{1309.7588}.
	The diagonal hatching
	of this bound in the bottom-left of the parameter space
	represents the region where the bound may weaken.
	Lines corresponding to nuclear and atomic density are shown for illustrative purposes.}
        \label{fig:secondfigure}

 \end{figure*}
\noindent The maximum area that gets melted or vaporized is then
\begin{equation}\label{max}
\pi r^2|_{max} = 4\pi t_{max}\alpha 
=\frac{\sigma_X v_X^2}{c_p T_c e}\,,
\end{equation}
which occurs at a time 
\begin{equation}\label{tmax}
t_{max} = \frac{\sigma_X v_X^2}{4 \pi \alpha c_p T_c e}\end{equation} 
after the macro passage. 

Note that this calculation ignores the specific heat of melting or vaporization, but we have checked that this is a small fraction of the deposited energy.

The expected number of macro passages through the rock
depends on $M_X$. 
\begin{align}
N_{events}&=\frac{\rho_{DM}A_{det}T_e v_X}{M_X} \nonumber \\
	&\approx 20 
	\left(\frac{A_{det}}{10m^2}\right)
	\left(\frac{T_e}{500My}\right)
	\left(\frac{kg}{M_X}\right)
	\,,
\end{align}
where 
{\color{black}$\rho_{DM} \approx 5.3 \times 10^{-25}\,$g$\,$cm$^{-3}$\cite{Bovy2012}},
$A_{det}$ is the cross-sectional area of the rock slab(s),
and $T_e$ is the exposure time, 
i.e. time the rock has spent near enough the surface
to be exposed to the high-velocity macro flux.

In Figure \ref{fig:rock}, we present 
the regions of parameter space that could be probed 
by three possible searches,
to highlight the different approaches
that could be taken. 
We imagine a process that takes thin slabs of rock
and inspects their surfaces for elliptical ``melt patches'' -- the cross-sections of the circular melt cylinder.
A macro would induce identical, aligned, elliptical
patches on the front and back sides of a thin
slab of uniform composition.
The hypothesis that a dark elliptical region
in a slab of light granite was caused by a macro
could thus be substantiated
by examining the obverse surface for
a dark patch of matching size, ellipticity and allignment.
There is no known alternate natural cause 
for such a pair of patches in granite.

Slabs of light-colored granite,
such as are commonly used for kitchen counter tops,
would seem to be ideal targets for inspection.
These could be examined at commercial showrooms after they were polished, but before they were cut to size and installed.
The uncut slabs are typically $2-3$cm thick, and several square meters in area.

Melted regions of such granite will tend to be much darker and thus easy to see in the polished surface.
They could then be confirmed by examining the back surface.

As the cross-sectional area of the macro-induced melt-tube 
is proportional to the geometric cross-section of the macro, $\sigma_X$,
the features caused by smaller macros are presumably harder to identify.
On the other hand, lower-mass macros are more abundant.
For a fixed (say nuclear) density this presents 
an obvious trade-off between feature size and feature abundance.
The grain sizes in granite are often at the $1\,$mm
scale or below.  The smaller the grains in a given slab, the smaller the minimum feature size that can confidently be searched for in that slab.
The rate of false detections can be determined from the rate of detections on the polished surface that are not matched on the back surface.


Granite is often very old. 
A typical value for the exposure time 
might be $100-500$ million years,
although much older rocks are widely known.
We adopt $T_e=500$ million years as a fiducial value, 
but the precise region of parameter space that can be probed will
depend on the provenance of the granite slabs.

In Figure 1, 
all three regions assume a minimum feature diameter of $1\,$mm.
(We assume circular features for simplicity.)
The first region with diagonal hatching 
assumed a slab area of just $10\,$m$^2$,
the second, unhatched region
assumed a slab of area $100\,$m$^2$ and 
The third region, with horizontal hatching, 
assumed a slab area of  $1000\,$m$^2$.

In Figure 2, we have shown the range of parameter space
that will be probed by examining a fixed area of granite but by placing various
requirements on the sized of the tracks left by a passing macro.
The region with diagonal hatching 
assumes
searches for features with diameter of $0.1\,$cm or more, 
the unhatched region is for a minimum feature diameter of $1\,$cm, and
the horizontal-hatched region assumes a minimum feature diameter of $10\,$cm.

It is likely that the granite slabs we will use for this
search will have somewhat different grain sizes and that
the region probed will be a convolution of the various regions corresponding
to different sized features.

The calculation presented above assumed that the macro moves through the rock at a velocity equal to its impact velocity on the surface.  
At sufficiently high $\sigma_X$, this will no longer be true.  
In Figure 1, we insist that the $\sigma_X$ be small enough for a macro that hits the Earth's surface at $250\,$km$\,$s$^{-1}$ to penetrate $5\,$km of rock without slowing to $100\,$km$\,$s$^{-1}$. 
If we were able to confidently identify rock samples that remained closer to the surface, or were older, we would be able to probe to larger values of $\sigma_X$ 
and $M_X$.
(However, we expect that at densities approaching atomic density, the macro, like a meteoroid, would disintegrate in the atmosphere or on impact with the Earth's surface.)
A full experimental analysis would integrate over the possible incidence angles of the macros and the velocity distribution of the dark matter.

We see that a 
manageable search for features that can easily be identified by eye, 
in a quantity of granite slabs such as are normally found at a typical commercial countertop showroom, 
will begin to probe unexplored regions of parameter space, but not down to nuclear density.
Since we do not know the detailed microphysics of macros, it is valuable to probe all open parameter values.
Moreover, this search would serve as an important proof-of-concept for scaling up to the large-scale effort that would be required to push  a search for macro dark matter down to the nuclear-density line.
\smallskip
\acknowledgements
This work was partially supported by Department
of Energy grant DE-SC0009946 
to the particle astrophysics theory group at CWRU.

\bibliographystyle{apsrev4-1}
\bibliography{sedimentary_cites}

\end{document}